\documentclass[12pt,preprint]{aastex}

\usepackage{lscape}
\usepackage{wasysym}
\usepackage{natbib}
\usepackage{longtable}

\shortauthors{Fischer et al.}
\shorttitle{AGN Inclinations}

\begin{document}

\title{Determining Inclinations of Active Galactic Nuclei Via Their Narrow-Line Region Kinematics - II. Correlation With Observed Properties \altaffilmark{1}}

\author{T.C. Fischer\altaffilmark{2},
D.M. Crenshaw\altaffilmark{2},
S.B. Kraemer\altaffilmark{3},
H.R. Schmitt\altaffilmark{4},
T.J. Turner\altaffilmark{5}}

\altaffiltext{1}{Based on observations made with the NASA/ESA Hubble Space 
Telescope, obtained at the Space Telescope Science Institute, which is 
operated by the Association of Universities for Research in Astronomy,
Inc. under NASA contract NAS 5-26555. These observations are associated 
with programs 11243, 11611, and 12212}

\altaffiltext{2}{Department of Physics and Astronomy, Georgia State 
University, Astronomy Offices, 25 Park Place, Suite 600,
Atlanta, GA 30303; fischer@chara.gsu.edu}

\altaffiltext{3}{Institute for Astrophysics and Computational Sciences,
Department of Physics, The Catholic University of America, Washington, DC
20064}

\altaffiltext{4}{Naval Research Laboratory, Washington, DC 20375}

\altaffiltext{5}{Department of Astronomy, University of Maryland, College Park, MD 20742}

\begin{abstract}

Active Galactic Nuclei (AGN) are axisymmetric systems to first order; their 
observed properties are likely strong functions of inclination with respect to 
our line of sight, yet the specific inclinations of all but a few AGN are 
generally unknown. By determining the inclinations and geometries of nearby 
Seyfert galaxies using the kinematics of their narrow-line regions (NLRs), and comparing them with 
observed properties, we find strong correlations between inclination and total hydrogen 
column density, infrared color, and H$\beta$ full-width at half maximum (FWHM). These 
correlations provide evidence that the orientation of AGN with respect to our line of 
sight affects how we perceive them, beyond the Seyfert 1/2 dichotomy. They can also be used to 
constrain 3D models of AGN components such as the broad-line region and torus. 
Additionally, we find weak correlations between AGN luminosity and several modeled 
NLR parameters, which suggests that the NLR geometry and kinematics are dependent 
to some degree on the AGN's radiation field.

\end{abstract}


~~~~~

\clearpage

\section{Introduction}
\label{sec1}

Seyfert galaxies represent a relatively moderate luminosity ($L_{bol} \approx 10^{43} - 10^{45}$ 
erg s$^{-1}$), nearby (z $\leq$ 0.1) subset of the overall collection of AGN, which exhibit a dichotomy 
of broadened and unbroadened permitted emission lines. This dichotomy has led the Seyfert class of AGN to 
be divided into two groups \citep{Kha74}, with Seyfert 1s exhibiting spectra containing broad (full width 
at half-maximum [FWHM] $\geq$ 1000 km s$^{-1}$) permitted lines, narrower (FWHM $\leq$ 1000 km s$^{-1}$) 
forbidden lines, and distinct, non-stellar optical and UV continua, and Seyfert 2s containing only narrow permitted 
and forbidden emission lines, with their optical and UV continua being dominated by the host galaxy. 

\citet{Ost78} suggested that Seyfert 1s and 2s were physically the same objects, but that the BLRs of 
Seyfert 2s are obscured along our line of sight (LOS). The strongest observational evidence for a unified 
model stems from work by \citet{Ant85}, who discovered that weak, polarized broad emission lines do exist 
in Seyfert 2 galaxies by analyzing the polarized spectrum of the Seyfert 2 NGC 1068. This observation led 
to the generation of a unified AGN model, with AGN being similar objects, obscured by a toroidal structure 
of gas and dust in Seyfert 2s \citep{Ant93}, viewed from different angles. Observing an AGN from near the 
axis of the torus would allow for the viewing of the central source, the BLR, and the NLR, and produce the 
spectrum of a Seyfert 1, while observing the same AGN from near the plane of the torus would leave only 
the NLR and a small amount of emission from the nuclear source visible, producing a Seyfert 2 spectrum. 
It is important to note that this is an oversimplified scenario, in which each AGN with a given luminosity 
is contained within an identical, smooth-density torus. In reality, the torus around each AGN likely varies 
in covering factor, or how much sky at the AGN center is covered by obscuring material, and has a clumpy 
composition that does not create a hard cut-off between observing a Type 1 or Type 2 AGN, but instead only 
reduces the probability to directly view the AGN as our line of sight strays further from the torus axis 
\citep{Eli12}. The NLR is then the radiation that has escaped the torus, forming a biconical structure 
likely governed by the torus covering factor, with an axis that is perpendicular to the plane of the torus. 

In our previous paper (\citealt{Fis13a}, henceforth Paper I), we measured radial velocities of [O~III] 
emission knots across NLRs for 47 Seyfert AGN observed with {\it Hubble Space Telescope} ({\it HST}) STIS 
G430L/M gratings. From our measurements, 12 AGN showed clear evidence for  kinematics dominated by biconical 
outflow, where knots accelerated out from the inner nucleus, reached a terminal velocity, and then decelerated 
back toward the AGN systemic velocity. In order to determine the geometry, including the inclination of 
their NLR outflows, and thus the inclination of their torus, (both of which generally unknown for all but 
a few AGN) a kinematic model was fit to the observed NLR kinematics using a simple, linear velocity law. 
Using available imaging and kinematic data for each AGN to create an initial geometry, we refined the model 
in an iterative process until a best fit to the kinematic data was obtained. Including 5 AGN previously 
modeled by our group (\citealt{Cre00a,Cre00b,Rui01,Das05,Das06,Cre10a,Fis10,Fis11}), we now have an initial 
sample of 17 AGN inclinations, from which we can begin to determine how their observed properties vary as 
a function of polar angle (i.e. the inclination of the torus) with respect to the accretion disk and/or 
torus axes, and identify correlations that probe the structure of the AGN components by comparing our 
kinematic model results with observed multiwavelength properties of these AGN.

Section 2 describes the parameters and how they were measured. Section 3 illustrates all relevant 
comparisons between modeled and observed AGN parameters. Sections 4 notes significant non-correlations. 
Section 5 contains discussion and final conclusions.

\section{Parameters}
\label{sec2}
 
\subsection{Kinematic Model Parameters}

Our kinematic models employed in Paper I, combined several basic input 
parameters to generate a 3D biconical NLR which simulated the deprojected outflow
velocities as a function of radius from the AGN. To review, the parameters and their definitions are 
as follows. Position angle ($P.A.$) is the angle between North and the bicone axis in the plane of the sky, 
measured in the eastward (counter-clockwise) direction. Inclination ($i$) of the bicone is measured out of 
the plane of the sky, such that $i$ = 0 places the bicone axis in the plane of the sky and $i$ = 90 places the bicone 
axis parallel to our line of sight. Inner and outer opening angles ($\theta_{min}$,$\theta_{max}$) 
are angles between the bicone axis and the inner and outer edges of the bicone model 
respectively. Our model includes an inner opening angle as previous NLR outflow studies have shown 
a lack of low velocity kinematic measurements near the peaks of the kinematic curves, which suggests 
that the NLR is often evacuated of [O~III] emission along its axis. Maximum velocity ($v_{max}$) is 
the boundary value set in the model velocity law 
$v = kr$, where velocity increases with radius until the maximum velocity is reached at the turnover 
radius ($r_t$) then decreases linearly to zero. Maximum height ($z_{max}$) is the distances 
from the nucleus to one end of the bicone, measured along the bicone axis. Model parameters are 
initially set to observed values via imaging and spectroscopy and altered in an iterative process 
until the model agrees with the kinematic data. All values are deprojected (i.e. corrected for inclination). 
Final parameters for each modeled AGN are listed in Table \ref{modelvals}.

\subsection{Physical Properties}

To determine how NLR geometry might be correlated with observable parameters of AGN, we gathered 
measurements from the literature, and in a few cases our own observations, of several observed AGN 
properties likely to be affected by viewing angle with respect to the NLR bicone axis and/or torus 
inclination, which we designate as polar angle (or torus inclination) $\theta$ $= 90^{\circ}-i$, 
where $i$ is the inclination of the bicone axis out of the plane of the sky. These properties include 
total hydrogen column density in the line of sight to the nucleus ($N_H$), mid-IR color defined by 
continuum flux ratios from {\it Spitzer} observations, and full-width at half-maximum (FWHM) of 
$H\beta$ emission from the broad-line region. We list the observed and physical properties and their 
uncertainties in Table 2.

Bolometric luminosities were obtained by integrating across all available spectral energy distribution 
(SED) flux points \citep{Woo02} or derived from [O~III] ($log(L_{Bol}) \approx log(L_{[OIII]})+log(3500)$; 
\citealt{Hec04,Hec05,Mel08}) or 2-10 keV X-ray \citep{Mag98} luminosities. Super-massive black hole 
(SMBH) masses of Seyfert 1 AGN were determined using reverberation mapping, where BLR size, estimated 
from the time lag between corresponding ionizing continuum and broad line emission fluxes, and 
broad-line velocities are combined to calculate the AGN black hole mass \citep{Pet04,Ben06,Den09,Den10}. 
SMBH masses for Seyfert 2s Circinus and NGC 1068 were gathered employing direct, dynamical 
mass-measurements  \citep{Gre03b,Lod03}. Remaining Seyfert 2 AGN SMBH masses were indirectly constrained 
utilizing stellar velocity dispersions as they correlate well with black holes masses determined from 
spatially resolved kinematics \citep{Tre02}. 

Type 2 AGN neutral column densities are measured directly through X-ray observations. As the 
column density of material, containing mostly hydrogen, within the obscuring torus 
surrounding the AGN increases, more ionizing radiation is absorbed. This effect can be 
prominently observed in the X-ray using the {\it Chandra}, {\it XMM-Newton}, and {\it Suzaku} 
observatories, where emission from an obscured AGN is absorbed below a specific energy due primarily 
to the H and He edges \citep{Tur09}. The emission that remains is fit with an absorbed power law model 
to determine $N_H$ \citep{Win09}. Values with lower limits are due to X-ray absorption extending to 
energies higher than could be observed by a specific observatory, which could therefore not be fit 
with a power law. In cases where the absorber is Compton thick, the lower limit is N$_{H} \approx$ 
10$^{24}$ to 10$^{25}$ cm$^{-2}$, depending on the observations. For Seyfert 2s, we assume that the total 
hydrogen column density in the line of sight to the nucleus is dominated by the neutral gas, i.e. N$_{H} 
\approx$ N$_{HI}$.

For Type 1 AGN, where we can assume that we are looking into the inner portion of the AGN (as verified 
in Paper I), we do not observe large neutral hydrogen column densities representative of the material 
found in the torus surrounding the central engine. Instead, we typically measure ionized column densities 
from ``AGN winds'' \citep{Cre03}, generally seen in the X-ray but have also been traced in the UV \citep{Cre12}, 
and determine $N_H$ from photoionization models. In short, intensities of two or more resonance doublet line species (e.g. 
C~IV, N~V) are measured which allow for the calculation of the covering factor of the outflowing absorbers 
with respect to the background emission, which in turn allows for the calculation of the optical depth of 
each absorption feature. The ionic column density of each absorption line is then obtained by integrating 
the optical depth across the line profile \citep{Cre03}. These ionic column densities are used to determine 
the ionization parameter (ratio of ionizing photon density to gas density, U) and $N_H$ (= $N_{HI} + N_{HII}$) 
via CLOUDY photoionization modeling \citep{Kra09,Kra12}. The listed $N_H$ values for each Sy 1 in our sample 
are the summed column densities across all absorption components. Uncertainties in the column densities for 
both Seyfert 1s and 2s are from the model uncertainties in the original references. For cases with multiple 
values, due to variability for example, we used an average value and standard deviation for the column 
densities and their uncertainties.

For the Seyfert 1 galaxies, we include the UV and X-ray``warm absorbers'' 
\citep{Cre12}, but not the highly-ionized ultra-fast outflow (UFO) absorbers detected in 
the Fe K-shell absorption lines, because the two types are quite distinct in 
terms of ionization parameter, outflow velocity, distance from the central AGN, 
and column density (\citealt{Tom13}, and references therein).  UFOs may therefore 
represent a  physically distinct component of AGN, whereas a strong connection has 
already been established between the UV and X-ray warm absorbers \citep{Cre03}. We 
will include the high-ionization absorbers (at all outflow velocities) in a future 
study to investigate their dependence on inclination and  possible connection to the warm absorbers.

\begin{figure}
\centering
\includegraphics[angle=0,scale=.8]{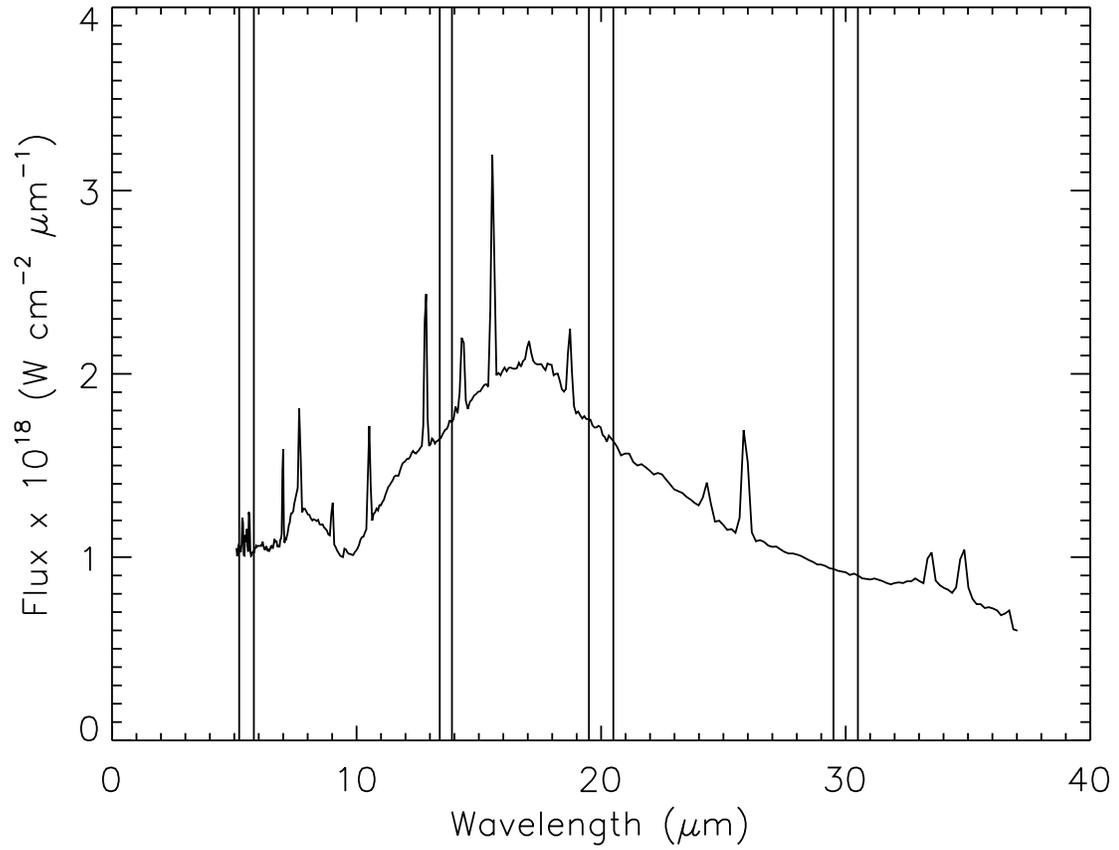}
\caption[Mid-IR Flux Measurement]{Mid-IR {\it Spitzer} spectrum of Mrk 3 gathered from CASSIS.
Vertical lines from left to right denote 5.5$\mu$m, 13.7$\mu$m, 20$\mu$m, and 30$\mu$m flux bins.
}
\label{color}
\end{figure}

We retrieved mid-IR spectra from the Cornell Atlas of Spitzer/IRS Sources (CASSIS) 
\citep{Leb11}, which provides reduced low-resolution spectra ($R\sim60-127$ over $5.2\mu$m to $38\mu$m) 
taken by the Infrared Spectrograph (IRS) instrument aboard {\it Spitzer}. Spectra were extracted through 
a dedicated pipeline from which we chose a regular ``tapered'' extraction suggested for extended sources.
Measuring continuum fluxes over small bins at 5.5$\mu$m, 13.7$\mu$m, 20$\mu$m, and 30$\mu$m (shown in 
Figure \ref{color}) allowed us to determine the mid-IR color of each AGN. Each observation was obtained 
in four segments that required scaling before each segment was combined into a single spectrum. Segments 
were scaled such that overlapping ends of each segment were equal in flux, creating a continuous spectrum 
when combined. As we are interested in relative fluxes, accurate absolute flux values are not a 
concern.

The FWHM of the broad component of H$\beta$ in each Seyfert 1 was obtained from the literature 
or our own observations. Observations of NGC 3227 were obtained using the DeVeny spectrograph on 
the 1.8m Perkins telescope at the Lowell Observatory in Flagstaff, Arizona. Spectra were taken on 
2010 December 9 using a slit width of $2''$ for a total exposure of 1200s and dispersed into spectral 
images using the 'blue' 300 line mm$^{-1}$ grating with a resolving power of $\sim$2145 ($\sim$140 
km s$^{-1}$).Observations for NGC 3783 was obtained using the R-C spectrograph on the 1.5 SMARTS 
telescope at the Cerro Tololo Inter-American Observatory (CTIO) in La Serena, Chile. Long-slit 
spectra were taken on 2003 April 25 for a total exposure time of 1800s using the 600 line mm$^{-1}$ 
'$\#$26' grating with a resolving power of $\sim$1165 ($\sim$257 km s$^{-1}$). Both sets of 
observations were reduced two-dimensionally using their respective IRAF packages and collapsed to a 
single spectrum, resulting in wavelength calibrated spectra in flux units from which FWHMs could be 
measured.FWHM measurements for the remaining targets were obtained from the literature from studies 
using similar techniques, which included the removal of the narrow component of H$\beta$. We estimate 
our uncertainties in FWHM by using different reasonable continuum placements and estimates for the 
narrow H$\beta$ fluxes, which typically result in percentage errors of 10\%.

Molecular hydrogen masses (M$_{H_2}$) were estimated via K-band IFU observations from SINFONI 
and OSIRIS by \citet{Mul09,Hic09}and \citet{Fri10}. In these studies, Gaussians were fit to the 
bright 1-0 S(1) molecular hydrogen line at 2.1218 $\mu$m across the inner 30 pc of each AGN to 
measure the rotational velocity of the system and calculate a dynamical mass within that radius, 
using the assumption that the gas is in a disk-like distribution around the AGN, from which a 
molecular hydrogen gas mass was estimated using a typical gas mass fraction of 10\%. Masses listed 
for Circinus and NGC 1068 are not included in our correlation analysis as the mass estimation for 
Circinus was only done for the inner 9 pc of the AGN and the mass in NGC 1068 may be miscalculated 
as the inflowing $H_2$ kinematics do not conform to the rotational pattern seen in other AGN. NGC 
3227 was observed by both listed instruments and thus the average between the two mass estimates 
is used.

\section{Correlations}
\label{sec3}

Using the kinematic modeling results from Paper I, we have the opportunity to compare geometrical 
aspects of the AGN NLRs to their other observed physical properties in order to detect any correlations 
between them for the first time. Table \ref{coefficients} lists the correlations examined, their 
resultant correlation coefficient $r$, the sample size $N$, and the probability (or p-value) of exceeding 
$r$ using a random sample of $N$ observations taken from an uncorrelated parent population $P_{c}(r,N)$. A 
p-value $< 5\%$ indicates a statistically significant correlation that is highly unlikely to be 
observed under a null hypothesis. The maximum sample size is one less than our total AGN sample size, as we 
omit NGC 5506 from our comparisons, as explained in Section 4.

\subsection{Inclination}

We can determine how observed AGN properties vary as a function of polar angle with respect to 
the accretion disk and/or torus axes by comparing them to the modeled inclinations in our AGN 
sample. The unified model of AGN posits the presence of a torus surrounding the AGN where the 
central engine and source of broad line emission is visible in Type 1 AGN and obstructs our view 
to the central engine in Type 2 targets. 

\begin{figure}
\centering
\includegraphics[angle=0,scale=0.8]{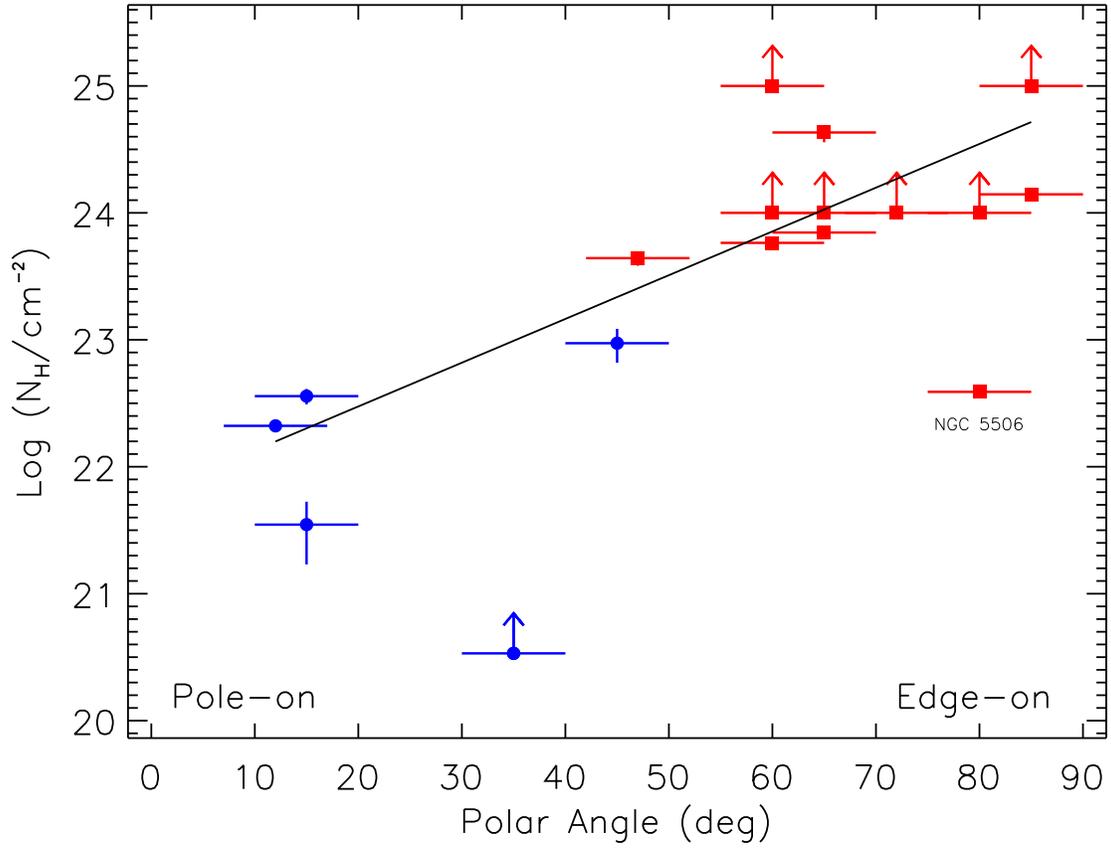}
\caption[Inclination vs. Column Density]{Inclination from our line of sight to the bicone axis versus 
the x-ray column density (ionized for Seyfert 1s, cold for Seyfert 2s), 
where a continuous trend can be seen between Seyfert 1s and 2s. Seyfert 1s are designated as 
blue circles, Seyfert 2s are designated as red squares. Points with arrows show lower 
limits of column densities.
}
\label{columns}
\end{figure}

Figure \ref{columns} compares column density N$_{H}$ against the polar angle $\theta$ ($= 90 - i$) 
of the bicone axis for each of our modeled AGN. Including both types of Seyferts, there is a distinct 
correlation between the two parameters, where observing an AGN further from the axis of its NLR and 
closer to the plane of the toroidal structure corresponds to a increased neutral hydrogen column 
density along our line of sight to the central engine. By removing NGC 5506 (an alteration kept for 
the remainder of this work), and Mrk 279, as column densities from several its absorbers have yet to 
be determined \citep{Cre12}, we calculate a coefficient of $r = 0.86$, which corresponds to a p-value 
of $P_{c}(r,N) < 0.01\%$, indicating that as our line of sight with respect to the bicone axis (polar 
angle) increases, the torus surrounding the AGN becomes closer to edge-on and we see an increase in 
column densities for both Seyfert 1s and Seyfert 2s. Seyfert 1s and 2s can reside at similar inclinations 
depending on whether or not our line of sight intersects the obscuring torus surrounding the AGN. 
Surprisingly, there appears to be a seamless transition in N$_{H}$ between the Seyfert 1 ionized gas 
columns and Seyfert 2 neutral columns. As we expected the total hydrogen column (N$_{H}$) in Seyfert 
2s would be dominated by neutral hydrogen and Seyfert 1s show large columns of ionized gas that 
dominate N$_{H}$ \citep{Cre99}, this correlations suggests a possible connection between the torus 
and the outflowing winds of ionized gas.

\citet{Deo09} analyzed mid-infrared (mid-IR) spectra of Seyfert galaxies using archival {\it 
Spitzer Space Telescope} observations in order to characterize the nature of the mid-IR active 
nuclear continuum, which is dominated by dust emission. They found that Seyfert 2s typically 
have weaker short-wavelength (5.5 - 14.7 $\mu$m) nuclear continua than comparable Seyfert 1s, 
which suggests a relationship between the mid-IR flux ratio ($F_{5{\mu}m}/F_{30{\mu}m}$)and 
inclination, as short-wavelength mid-IR emission is likely absorbed to a certain degree in Type 
2 objects by the obstructing torus surrounding the AGN. According to the unified model, as our 
view becomes more pole-on, the 5 $\mu$m emission should increase as we see more of the hot, 
inner throat of the torus. Comparing our results with \citet{Deo09} for targets that are 
contained in both their sample and CASSIS (Mrk 3, NGC 4151, NGC 4507), we find our measured 
fluxes to be consistent to within 10\%. As observations of certain sample AGN were not 
available in CASSIS, we then use flux ratios from \citet{Deo09} for  NGC 1667, NGC 4051, and 
NGC 7674.

\begin{figure}
\centering
\includegraphics[angle=0,scale=0.8]{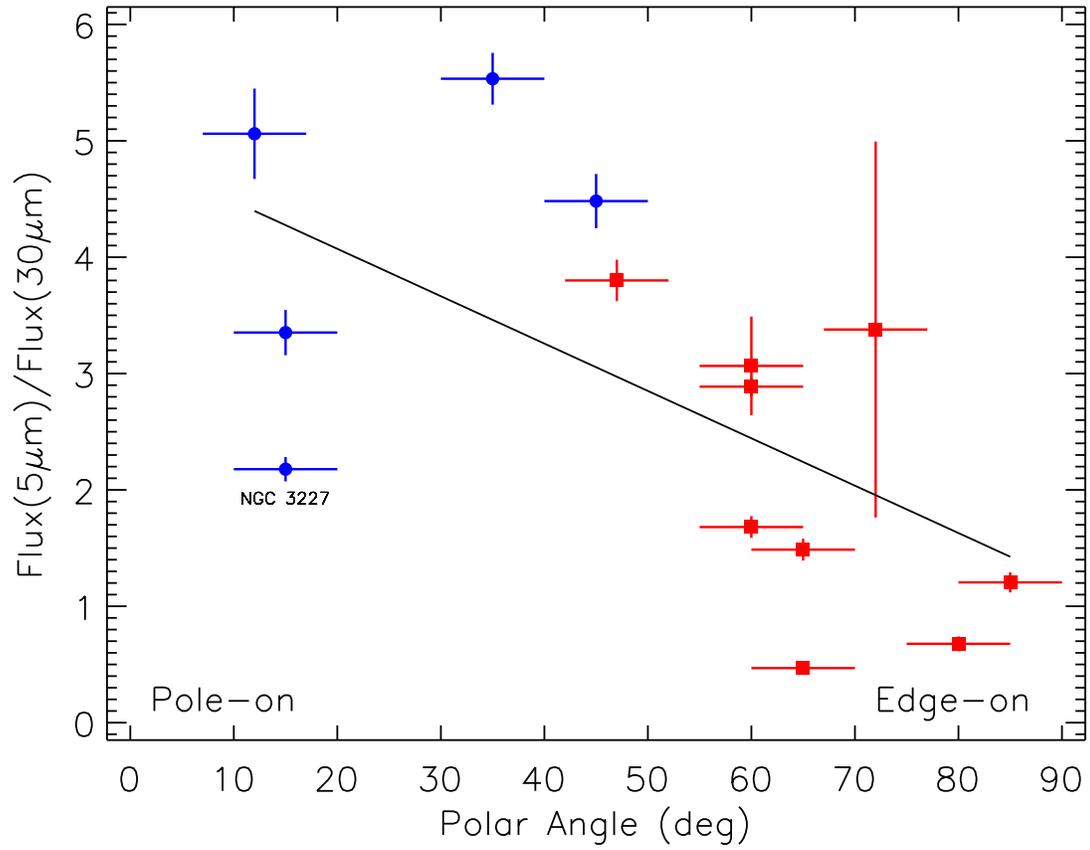}
\caption[Inclination vs. Mid-IR Color]{Inclination from our line of sight to the bicone axis versus 
AGN mid-IR color ($F_{5.5{\mu}m}/F_{30{\mu}m}$). Seyfert 1s are designated as 
blue circles, Seyfert 2s are designated as red squares.
}
\label{colors}
\end{figure}

Figure \ref{colors} shows a negative correlation between mid-IR color and polar angle with a 
correlation coefficient of $r = -0.61$ $(P_{c}(r,N) = 1.2\%)$, supporting the idea that a smaller polar angle 
allows for a better view of the hot inner 'throat' of the obscuring torus. Removing NGC 3227, a 
Seyfert 1 galaxy known to be heavily reddened by its host galaxy \citep{Cre01}, increases the 
correlation to $r = -0.74$ $(P_{c}(r,N) = 0.16\%)$.

\begin{figure}
\centering
\includegraphics[angle=0,scale=0.8]{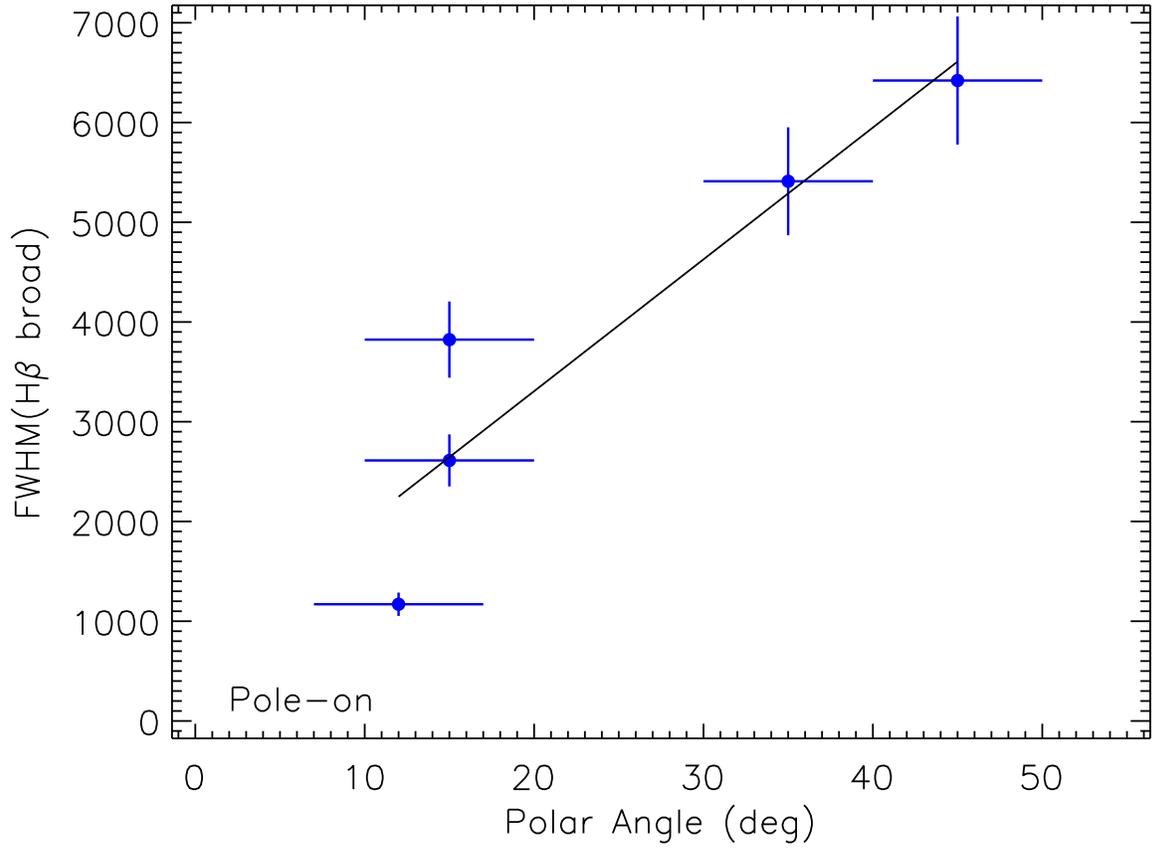}
\caption[Inclination vs. H$\beta$ FWHM]{Inclination from our line of sight to the bicone axis versus 
H$\beta$ FWHM for modeled Seyfert 1s.
}
\label{fwhm}
\end{figure}

By observing the FWHM of broad H$\beta$ emission-lines in Sy 1 targets and comparing them with 
their modeled inclinations, we may gain a better understanding of the BLR kinematics. Figure 
\ref{fwhm} shows a positive correlation ($r = 0.94, P_{c}(r,N) = 1.7\%$) between broad H$\beta$ 
FWHM and polar angle, despite the small number of data points, which suggests a non-spherical 
component to the BLR kinematics. If the BLR kinematics contain a strong rotational component, 
as suggested by a number of studies \citep{Mur97,Gas00,Gas09}, this correlation suggests that 
pole-on and near pole-on AGN contain velocities that may be underestimated. Thus, using the mass 
relation via reverberation mapping \citep{Pet97}, $M = f\frac{R{\Delta}V^{2}}{G}$ where $R$ is 
the size of the BLR, ${\Delta}V$ is the emission line width, and $f$ is a scale factor of order 
unity dependent on the inclination of the BLR, amongst other things, the correlation also 
suggests that black-hole masses may be underestimated in some AGN \citep{Col06}. Although this 
correlation is suggestive, it cannot be used directly because the FWHM is also dependent on 
black hole mass and size of the BLR.

\subsection{Luminosity}

\begin{figure}
\centering
\includegraphics[angle=0,scale=0.8]{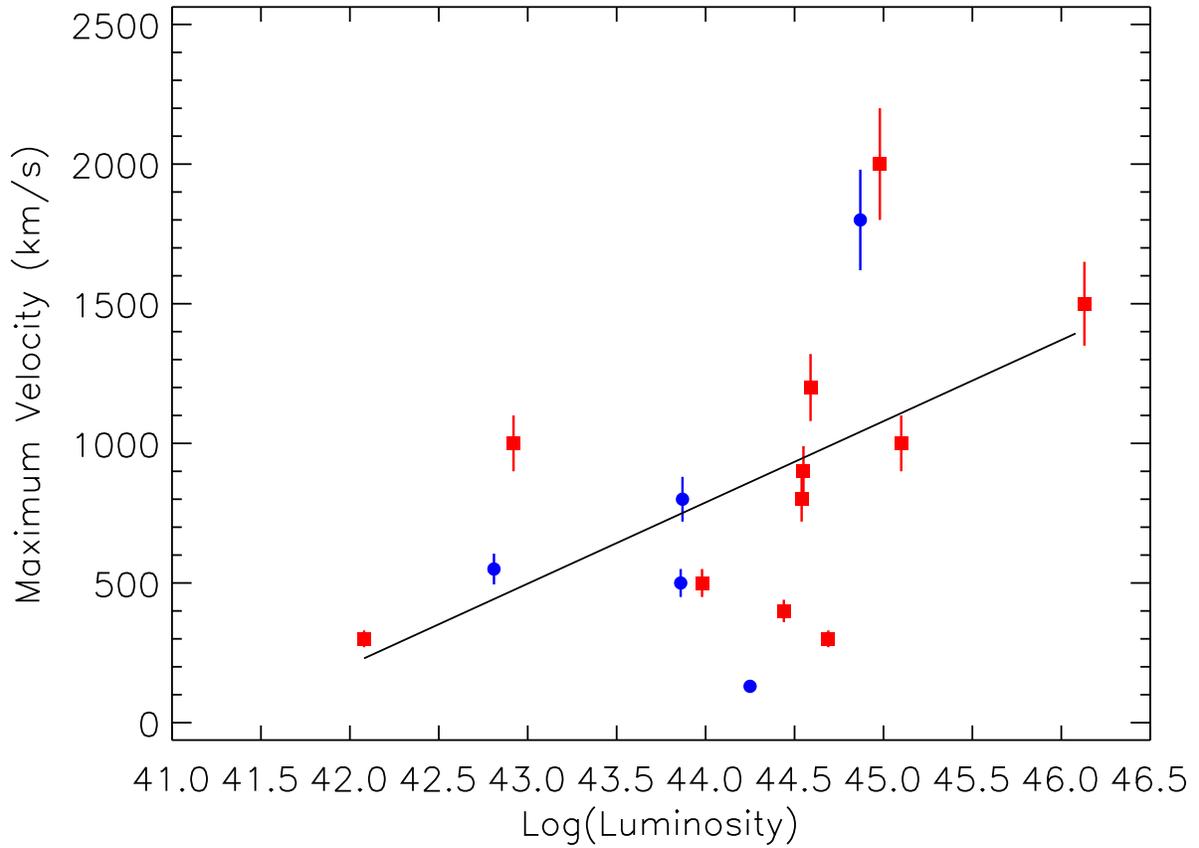}
\caption[Bolometric Luminosity vs Maximum Velocity]{Bolometric luminosity versus maximum 
deprojected outflow velocity of the NLR for each of our modeled targets. Seyfert 1s are designated as 
blue circles, Seyfert 2s are designated as red squares. Luminosity are uncertain by approximately a factor of two.
Modeled parameters contain uncertainties of approximately 10\% the modeled value.
}
\label{lum3}
\end{figure}

\begin{figure}
\centering
\includegraphics[angle=0,scale=0.8]{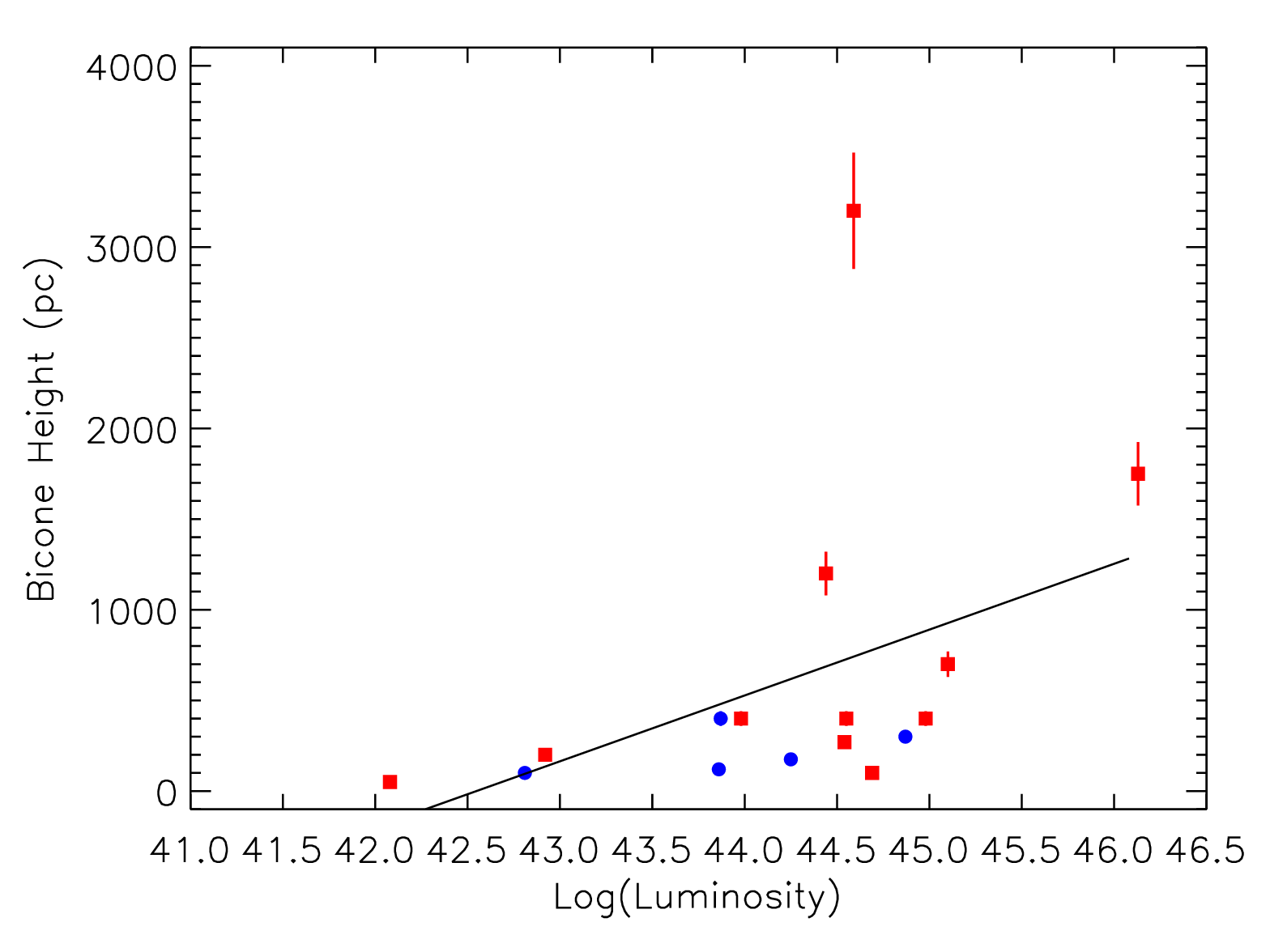}
\caption[Bolometric Luminosity vs. Bicone Height]{Bolometric luminosity versus NLR bicone height 
for each of our modeled targets. Seyfert 1s are designated as blue circles, Seyfert 2s are 
designated as red squares. Luminosity are uncertain by approximately a factor of two. 
Modeled parameters contain uncertainties of approximately 10\% the modeled value.
}
\label{lum4}
\end{figure}

\begin{figure}
\centering
\includegraphics[angle=0,scale=0.8]{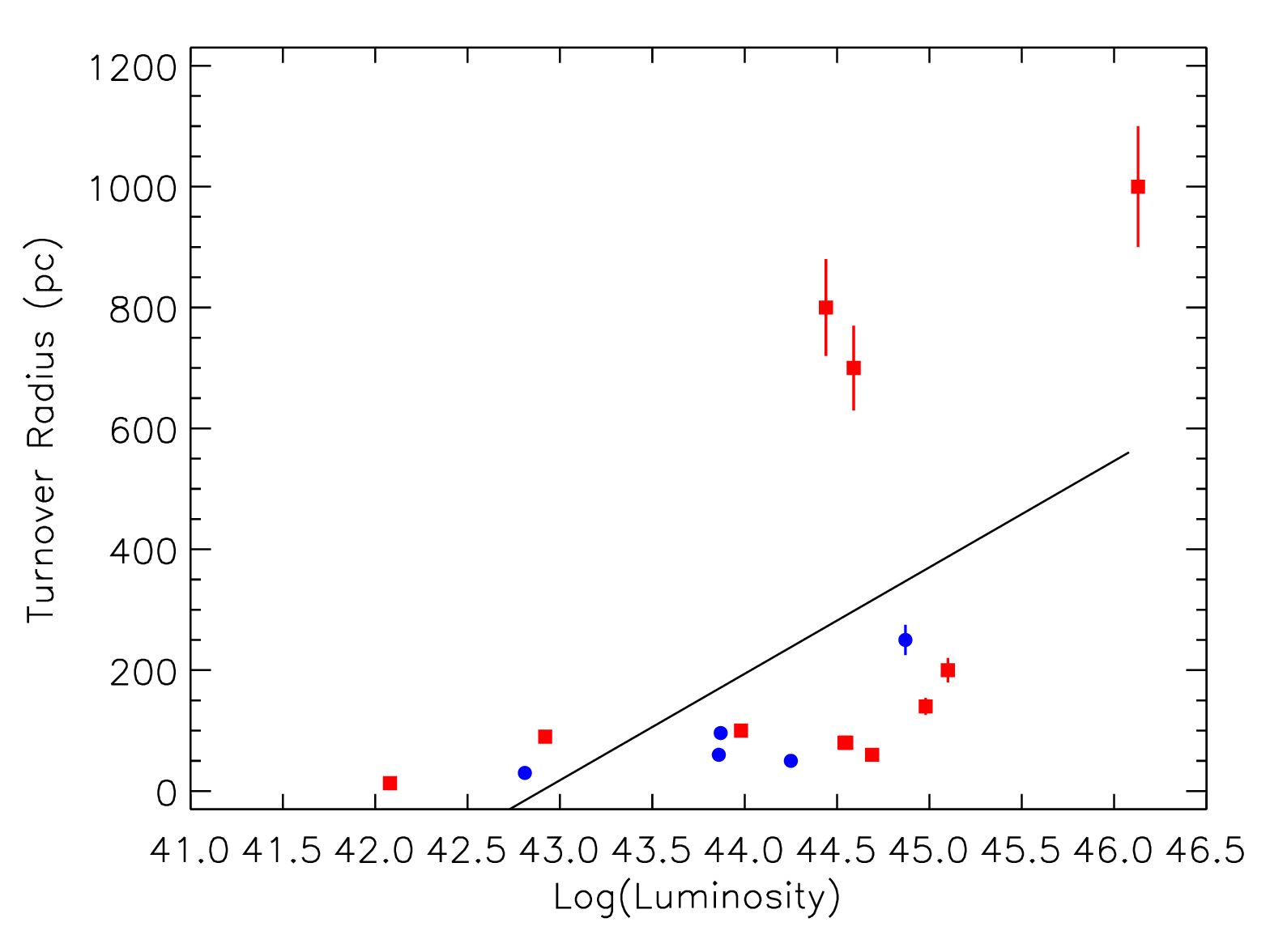}
\caption[Bolometric Luminosity vs. Turnover Radius]{Bolometric Luminosity versus kinematic turnover radius 
for each of our modeled targets. Seyfert 1s are designated as blue circles, Seyfert 2s are 
designated as red squares. Luminosity are uncertain by approximately a factor of two. 
Modeled parameters contain uncertainties of approximately 10\% the modeled value.
}
\label{lum5}
\end{figure}

In addition to several observed parameters correlating with AGN inclination, we also found that 
several {\it model} parameters, $r_t$, $v_{max}$, and $z_{max}$, each exhibit a possible positive 
correlation with AGN bolometric luminosity. Figures \ref{lum3}-\ref{lum5} suggest that as the 
AGN increases in luminosity, the maximum values of these parameters also increase. Correlations 
of v$_{max}$ and r$_t$ with luminosity suggest that radiative driving is an important factor 
within the NLR. As luminosity increases, the increased number of photons accelerate gas clouds 
in the NLR to higher velocities out to further distances before decelerating. The increased 
number of photons also continue to ionize gas further from the nucleus, increasing the total 
NLR bicone height. However, the correlations provided are very weak, so much so that comparing 
$z_{max}$ with luminosity results in a p-value of $P_{c}(r,N) > 5\%$, which indicates that the 
correlation is not statistically significant. Modeled NLR geometries for additional AGN are 
required before these relations can be further analyzed. 

\begin{figure}
\centering
\includegraphics[angle=0,scale=0.8]{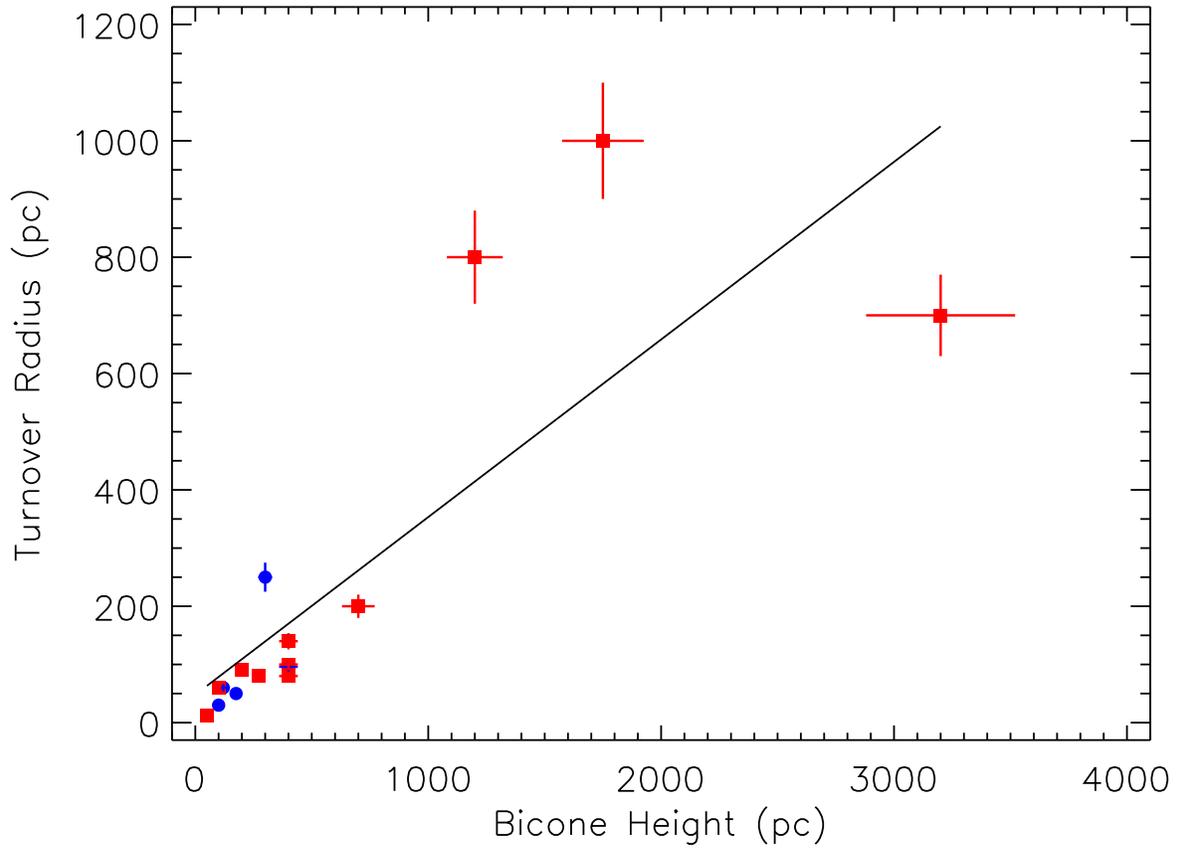}
\caption[Bicone Height vs. Turnover Radius]{NLR bicone height versus kinematic turnover radius 
for each of our modeled targets. Seyfert 1s are designated as blue circles, Seyfert 2s are 
designated as red squares. Modeled parameters contain uncertainties of approximately 10\% the modeled value.
}
\label{lum6}
\end{figure}

Additionally, we find a strong  ($P_{c}(r,N) = 0.01\%$) correlation between bicone height and 
kinematic turnover radius (Figure \ref{lum6}), which suggests that geometric and kinematic 
scales are related, likely via AGN luminosity. 

\section{Non-Correlations}

Through the Unified Model, the central engine is obstructed by an optically thick disk \citep{Ant93} 
such that a majority of the ionizing radiation is collimated into a biconical structure that has been 
observed to vary in opening angle amongst different AGN \citep{Sch03}. As such, any correlations including 
the NLR opening angle may provide information about the composition of the surrounding torus. However, 
of all the intrinsic NLR model parameters (i.e. not inclination or position angle), outer opening angle 
appears to be highly uncorrelated with both observed and modeled AGN properties in our sample, as shown 
by the high $P_{c}(r,N)$ values for correlations with $\theta_{max}$ in Table \ref{coefficients}. These results disagree 
with previous correlations found by \citet{Mul11} using similar NLR modeling techniques, where opening 
angle, maximum outflow velocity, and molecular hydrogen mass within the inner torus, 
each appeared dependent on one another, indicating that more massive tori restrict NLRs from expanding 
in opening angle and produce higher velocity outflows. Some of the discrepancy may be that we do not 
have enough dynamic range as our molecular mass comparisons contain a very small sample size of four AGN. 
However, we are able to compare outflow velocity and opening angle for our entire sample and find little 
correlation between these two parameters.

\section{Discussion and Conclusions} 
\label{sec4}

With inclinations and geometries of 17 Seyfert galaxies, our model results suggest that 
Seyfert 1 AGN are inclined further toward our LOS than Seyfert 2 AGN. Knowing the inclinations of these AGN 
allows us to determine how their observed properties vary as a function of polar angle with respect to 
the accretion disk and/or torus axes. We have established a strong connection between polar angle and column density, 
which probes the structure of the obscuring torus, the ionized gas outflows, and the connection between 
the two; polar angle and broad H$\beta$ FWHM, suggesting a non-spherical component to the BLR kinematics; 
and polar angle and mid-IR color (5 $\mu$m/30 $\mu$m flux), presumably indicating a better view of the hot 
inner ``throat'' of the obscuring torus in Seyfert 1s compared to Seyfert 2s.

With the discovery of three independent correlations on inclination, we can see that 
our line of sight affects how we observe specific parameters of our sample and 
thus that the obstructing torus surrounding the central engine plays a 
significant role in how we observe these AGN. Additionally, we find correlations 
between AGN luminosity and the NLR kinematics and geometry, but not between opening angle, 
molecular hydrogen mass surrounding the torus and the NLR kinematics and geometry.

Our inclination results agree with and help quantify the unified model, as $N_H$, IR color, and $H\beta_{FWHM}$ 
measurements each show a strong correlation with viewing angle. These results can thus provide valuable constraints 
on specific torus models (e.g. \citealt{Fri06,Ram09,Rot12,Sta12}) and BLR models (e.g. \citealt{Col06,Den12,Pan12}). 

Establishing accurate, continuous correlations between AGN inclination and independent, observable parameters may allow us 
to use these comparatively simple to obtain observational measurements in the future as a proxy for 
inclination in AGN with unmodelable kinematics. What other observable parameters could be related to 
inclination? Could the amount of polarized light depend on inclination? As \citet{Ant93} observed, 
polarization of type 1 galaxies tends to be parallel to the AGN jet axis and polarization of type 2 
galaxies tends to be perpendicular to the jet axis. Could this also apply to the torus axis, such that 
a correlation between increasing perpendicular polarization and polar angle exists? 

One parameter that certainly requires further investigation is the role of the host galaxy in obscuring the 
AGN. As mentioned earlier, the maximum sample size is one less than our total AGN sample size, 
as we omit NGC 5506\footnote{Paper I stated that the classification of 
NGC 5506 was contested, with labels of Narrow-Line Seyfert 1 (NLS1;\citealt{Nag02}) / Sy 1.9 
\citep{Mai95} / Sy 2 AGN \citep{Tri10} all being used to describe the AGN. However, \citet{Tri10} notes 
that the alternate labels do not correspond to what is seen optically in NGC 5506 and we have removed the 
other labels from contention in our analysis.} from our comparisons because it contains a highly inclined 
(76$^{\circ}$) host disk. Imaging depicts the NLR as a single cone near perpendicular to the host disk, 
and in Paper I, we modeled the polar angle of the NLR to be 80$^{\circ}$ from LOS. However, it is unknown 
how much of the visible NLR is extinguished due to the host disk orientation, although X-ray observations 
of NGC 5506 have shown that at least a portion of the NLR is extinguished as X-ray emission from a second NLR cone 
can be clearly detected on the opposite side of the nucleus from the visible NLR \citep{Zen09}. As we found 
no correlation between NLR and host disk orientation in Paper I, it is possible that the near edge-on host 
disk would play a major role in obscuring portions of a moderately inclined NLR, and the interpretation of 
only the unobscured kinematics would thus affect the resultant kinematic model. If the NLR were to be 
inclined only 40 degrees from our LOS, per Fe K$\alpha$ analysis in \citet{Gua10}, then the NLR may be 
opened wide enough that a portion of the NLR intersects with the disk. Several other modeled AGN experience 
intersections between the host and NLR (i.e. Mrk 3, Mrk 34, Mrk 573) which illustrate that the NLR is 
radiation bound in an intersection scenario and emission near this intersection does not continue to radially 
progress through the host disk. As such, an observer near the edge of the host disk of each of these AGN would 
likely be unable to observe an integral portion of the NLR kinematics. Additionally, while host disks likely 
play a role in the amount of NLR emission that is extinguished in each AGN, they do not contain hydrogen 
column densities significant enough to account for the X-ray/UV absorption observed over each AGN.  
Comparing host disk inclination to hydrogen column density in our sample results in a correlation coefficient of 
$r = 0.034$ and a corresponding p-value of $P_{c}(r,N) = 89\%$. Thus, at its current NLR orientation, NGC 5506 
has a hydrogen column density orders of magnitude lower than all other AGN at high inclinations, implying 
that it should have an inclination closer to our LOS. A polar angle of $\theta = 40^{\circ}$ would place the 
$N_{H} = 2 \times 10^{22}$ column density measurement much closer to the given correlation.

Within our limited range of luminosities, we find that outflow velocity, NLR height, and turnover radius 
each appear to be influenced by the luminosity of the AGN. Additionally, NLR 
height and turnover radius are well correlated, which means whatever is determining the ultimate reach 
of photons from the AGN also has some role in the dynamics of the system, extending the acceleration of 
NLR clouds to further distances. Could these relations provide insight on the origin of the NLR? If the 
NLR emitting gas primarily originates from a stream of ionizing radiation unleashed upon the galaxy, 
ionizing and accelerating ambient material that it encounters, the one property that should affect the 
geometry of the NLR the greatest would be AGN luminosity. To more properly understand this relation requires 
additional models to further constrain the strength of the given correlations and further study of the 
orientation between the NLR radiation and the plane of the host galaxy. AGN that display 
such an intersection (i.e. Mrk 78, NGC 4151) typically result in a more extended NLR and are more likely to 
host in situ acceleration of NLR gas \citep{Fis10}.

From our relation between bolometric luminosity 
and maximum outflow velocity, we see that outflow kinematics possibly cease at L$_{bol} < 10^{41}$ erg 
s$^{-1}$. {\it HST} STIS H$\alpha$ studies similar to our own in \citet{Wal08} depict observed low-ionization 
nuclear emission-line regions (LINERs), AGN with luminosities generally lower than Seyferts, containing 
possible rotation (or 'ambiguous'; see Paper I) kinematics with blueshifted velocities $<$ 250 km s$^{-1}$. Though 
these velocities may be due to projected outflows, they could instead be a combination of rotation and 
ionization via star formation or post-AGB stars \citep{Yan12}.

Finally, we note that although the correlations that we found are intriguing, they are based on a small number of 
data points. More confirmations of radial outflows in AGN and corresponding kinematic models would result in more 
inclinations for testing these correlations. More observations of Seyfert 2s at higher X-ray energies would 
help replace lower limits on $N_H$ with actual values and additional X-ray observations of our current sample 
over several epochs would allow us to determine if and how much $N_H$ varies over time.

\section{Acknowledgments}

TCF thanks M.C. Bentz, H.R. Miller, R.J. White, and P.J. Wiita for the useful discussions and A. Michel for 
aiding in H$\beta$ FWHM measurements. The authors would also like to thank the anonymous referee for their 
practical suggestions which led to a much improved discussion. Some of the data 
presented in this paper were obtained from the Mikulski Archive for Space Telescopes (MAST). STScI is 
operated by the Association of Universities for Research in Astronomy, Inc., under NASA contract NAS5-26555. 
This research has also made use of the NASA/IPAC Extragalactic Database (NED) which is operated by the Jet 
Propulsion Laboratory, California Institute of Technology, under contract with the National Aeronautics and 
Space Administration.

\clearpage

%TABLES%%%%%%%%%%%%%%%%%%%%%%%%%%%%%%%%

\begin{deluxetable}{l|rlccrrr|rcc|r}
\tablecolumns{11}
\tablecaption{Total Sample Modeled AGN Parameters.\tablenotemark{1}}
\tablewidth{0pt}
\tablehead{
\multicolumn{1}{c|}{Target}         & 
\multicolumn{7}{c|}{NLR Bicone}     & 
\multicolumn{3}{c|}{Host Disk}      \\
\multicolumn{1}{c|}{}               & 
\multicolumn{1}{c}{$P.A.$}          &      
\multicolumn{1}{c}{$i$~\tablenotemark{2}}             & 
\multicolumn{1}{c}{$\theta_{min}$}  & 
\multicolumn{1}{c}{$\theta_{max}$}  & 
\multicolumn{1}{c}{$v_{max}$}       &  
\multicolumn{1}{c}{$z_{max}$}       & 
\multicolumn{1}{c|}{$r_{t}$}        & 
\multicolumn{1}{c}{$P.A.$}          &   
\multicolumn{1}{c}{$i$~\tablenotemark{3}}             & 
\multicolumn{1}{c|}{Disk}           &
\multicolumn{1}{c}{$\beta$~\tablenotemark{4}}         \\
\multicolumn{1}{c|}{}               &  
\multicolumn{1}{c}{($^{\circ}$)}    &
\multicolumn{1}{c}{($^{\circ}$)}    &
\multicolumn{1}{c}{($^{\circ}$)}    &  
\multicolumn{1}{c}{($^{\circ}$)}    &
\multicolumn{1}{c}{(km/s)}          &   
\multicolumn{1}{c}{(pc)}            &   
\multicolumn{1}{c|}{(pc)}           &  
\multicolumn{1}{c}{($^{\circ}$)}    &  
\multicolumn{1}{c}{($^{\circ}$)}    & 
\multicolumn{1}{c|}{Ref.}           &
\multicolumn{1}{c}{($^{\circ}$)}

}
\startdata
Circinus    &   -52	  & 25 (NW) &  36	   &   41  &  300 &     35  &      9  &         30	&    65      & 1   &  7\\
Mrk 3       &    89       & 05 (NE) &  ---         &   51  &  800 &    270  &     80  &        129	&    64      & 9   & 52\\
Mrk 34	    &   -32	  & 25 (SE) &  30	   &   40  & 1500 &   1750  &   1000  &         65	&    30      & 2   & 85\\
Mrk 78	    &    65	  & 30 (SW) &  10	   &   35  & 1200 &   3200  &    700  &         84 	&    55      & 2   & 87\\
Mrk 279	    &   -24	  & 55 (SE) &  59	   &   62  & 1800 &    300  &    250  &         33 	&    56      & 3,4 & 86\\
Mrk 573	    &   -36	  & 30 (NW) &  51          &   53  &  400 &   1200  &    800  &        103      &    30      & 3,2 & 44\\ 
Mrk 1066    &   -41	  & 10 (NW) &  15	   &   25  &  900 &    400  &     80  &         90 	&    54      & 3   & 45\\
NGC 1068    &    30	  & 05 (NE) &  20	   &   40  & 2000 &    400  &    140  &        115 	&    40      & 8   & 45\\
NGC 1667    &    55	  & 18 (NW) &  45	   &   58  &  300 &    100  &     60  &          5      &    39      & 4   & 46\\ 
NGC 3227    &    30	  & 75 (SW) &  40	   &   55  &  500 &    200  &    100  &        -31 	&    63      & 2,5 & 76\\
NGC 3783    &   -20	  & 75 (SE) &  45	   &   55  &  130 &    110  &     32  &        -15 	&    35      & 6   & 38\\
NGC 4051    &    80       & 78 (NE) &  10	   &   25  &  550 &    175  &     52  &         50 	&    05      & 6   & 15\\
NGC 4151    &    60	  & 45 (SW) &  15	   &   33  &  800 &    400  &     96  &         33	&    20      & 7   & 39\\
NGC 4507    &   -37	  & 43 (NW) &  30	   &   50  & 1000 &    200  &     90  &         65 	&    28      & 2   & 12\\
NGC 5506    &    22	  & 10 (SW) &  10	   &   40  &  550 &    220  &     65  &        -89      &    76      & 3   & 32\\
NGC 5643    &    80	  & 25 (SE) &  50	   &   55  &  500 &    285  &     70  &        136 	&    30      & 3   & 42\\
NGC 7674    &   -63       & 30 (NW) &  35	   &   40  & 1000 &    700  &    200  &         76 	&    40      & 2   & 42
\enddata
\tablenotetext{1}{Inclination direction (e.g. NW, SE, etc.) specifies which end of the NLR bicone is inclined out of the plane of the sky toward Earth}
\tablenotetext{2}{Inclination of 0$^{\circ}$ corresponds to an edge-on orientation}
\tablenotetext{3}{Inclination of 0$^{\circ}$ corresponds to an face-on orientation}
\tablenotetext{4}{Angle between the NLR bicone axis and the normal to the host galaxy disk\\
{\bf References}: (1) \citet{Fre77}, (2) \citet{Sch00}, (3) \citet{Kin00}, (4) NED, (5) \citet{Xil02}, (6) \citet{Hic09}, (7) \citet{Das05}, (8) \citet{Das06}, (9) \citet{Cre10b}}
\label{modelvals}
\end{deluxetable}

\begin{landscape}
\addtolength{\tabcolsep}{-4pt}
\begin{deluxetable}{lccccccccccccccc}
\tabletypesize{\scriptsize}
\tablecolumns{2}
\tablecaption{Physical properties of modeled AGN sample}
\tablewidth{0pt}
\tablehead{\colhead{Target}  & \colhead{Type} & \colhead{log(L$_{Bol}$)} & \colhead{Ref.}  & \colhead{log(M$_{BH}$)}         & \colhead{Ref.} & L/L$_{edd}$ & \colhead{N$_H$}                      & \colhead{Ref.} & \colhead{F$_{5.5{\mu}m}$}             &\colhead{F$_{30{\mu}m}$}             & \colhead{Ref.} & \colhead{H$\beta$ FWHM}\tablenotemark{1}  & \colhead{Ref.}& \colhead{log(M$_{H_2}$)} & \colhead{Ref.}\\
                             &                & \colhead{erg s$^{-1}$}   &                 & \colhead{M$_{\astrosun}$}&                &             & \colhead{$\times 10^{22}$ cm$^{-2}$} &                & \colhead{(W cm$^{-2}$ $\mu$m$^{-1}$)} &\colhead{(W cm$^{-2}$ $\mu$m$^{-1}$)} &                & \colhead{km s$^{-1}$}    &         & \colhead{M$_{\astrosun}$}  &          }
\startdata
Circinus & 2 &  42.08 & 1 & 6.23 & 5	& 0.01 & 430$^{+40}_{-70}$      & 13 & -----                          &-----                          & -----& ---  & ----- & 6.30 & 30   \\
Mrk 3    & 2 &  44.54 & 2 & 8.65 & 2	& 0.01 & 136$^{+3}_{-4}$        & 14 & 1.08$\pm0.08 \times 10^{-18}$  &8.96$\pm0.11 \times 10^{-19}$  & 26   & ---  & ----- & ---  & -----\\
Mrk 34   & 2 &  46.13 & 3 & ---  & --   & ---  & $>$100                 & 15 & 2.75$\pm0.17 \times 10^{-19}$  &1.85$\pm0.02 \times 10^{-19}$  & 26   & ---  & ----- & ---  & -----\\
Mrk 78   & 2 &  44.59 & 2 & 7.87 & 2	& 0.04 & 57.5$\pm5.8$           & 16 & 3.53$\pm0.19 \times 10^{-19}$  &2.10$\pm0.02 \times 10^{-19}$  & 26   & ---  & ----- & ---  & -----\\
Mrk 279  & 1 &  45.04 & 3 & 7.54 & 6	& 0.25 & $>$0.034               & 17 & 7.47$\pm0.26 \times 10^{-19}$  &1.35$\pm0.03 \times 10^{-19}$  & 26   & 5411 & 28    & ---  & -----\\
Mrk 573  & 2 &  44.44 & 2 & 7.28 & 2	& 0.11 & $>$100                 & 18 & 7.45$\pm0.18 \times 10^{-19}$  &2.58$\pm0.04 \times 10^{-19}$  & 26   & ---  & ----- & ---  & -----\\
Mrk 1066 & 2 &  44.55 & 2 & 7.01 & 2	& 0.27 & $>$100                 & 19 & 7.45$\pm0.70 \times 10^{-19}$  &1.10$\pm0.007 \times 10^{-18}$ & 26   & ---  & ----- & ---  & -----\\
NGC 1068 & 2 &  44.98 & 2 & 6.93 & 7	& 0.44 & $>$1000                & 20 & -----                          &-----                          & -----& ---  & ----- & 7.36 & 31   \\
NGC 1667 & 2 &  44.69 & 2 & 7.88 & 2	& 0.05 & $>$100                 & 21 & 7.50$\pm3.50 \times 10^{-19}$  &2.22$\pm0.06 \times 10^{-19}$  & 27   & ---  & ----- & ---  & -----\\
NGC 3227 & 1 &  43.86 & 2 & 6.88 & 8	& 0.01 & 0.35$\pm0.18$          & 22 & 1.42$\pm0.07 \times 10^{-18}$  &6.52$\pm0.07 \times 10^{-19}$  & 26   & 3823 & 26    & 7.31 & 32   \\
NGC 3783 & 1 &  44.59 & 2 & 7.47 & 6	& 0.05 & 3.6$\pm0.5$            & 17 & 2.37$\pm0.13 \times 10^{-18}$  &7.07$\pm0.13 \times 10^{-19}$  & 26   & 2612 & 26    & 6.47 & 32   \\
NGC 4051 & 1 &  43.56 & 2 & 6.24 & 9	& 0.17 & 2.1$\pm1.1$            & 17 & 2.09$\pm0.16 \times 10^{-18}$  &4.13$\pm0.04 \times 10^{-19}$  & 27   & 1170 & 29    & 6.70 & 32   \\
NGC 4151 & 1 &  43.73 & 2 & 7.66 & 10	& 0.03 & 9.4$\pm2.8$            & 17 & 6.23$\pm0.30 \times 10^{-18}$  &1.39$\pm0.03 \times 10^{-18}$  & 26   & 6421 & 28    & 7.15 & 32   \\
NGC 4507 & 2 &  42.92 & 1 & 6.13 & 11	& 0.05 & 43.9$^{+5.4}_{-5.7}$   & 23 & 2.09$\pm0.10 \times 10^{-18}$  &5.50$\pm0.05 \times 10^{-19}$  & 26   & ---  & ----- & ---  & -----\\
NGC 5506 & 2 &  44.05 & 1 & 6.88 & 11	& 0.12 & 3.7$\pm0.8$            & 24 & -----                          &-----                          & -----& ---  & ----- & ---  & -----\\
NGC 5643 & 2 &  43.98 & 1 & 6.79 & 11	& 0.12 & 70.7$^{+30}_{-10}$     & 25 & 4.73$\pm0.46 \times 10^{-19}$  &1.01$\pm0.004 \times 10^{-18}$ & 26   & ---  & ----- & ---  & -----\\
NGC 7674 & 2 &  45.00 & 4 & 7.58 & 12	& 0.27 & $>$1000                & 21 & 1.87$\pm0.26 \times 10^{-18}$  &6.10$\pm0.05 \times 10^{-19}$  & 27   & ---  & ----- & ---  & -----
\enddata
\tablenotetext{1}{H$\beta$ FWHM values gathered for Seyfert 1s only.\\
	  {\bf References}: (1) \citealt{Mel08}, (2) \citealt{Woo02} and references therein, (3) \citealt{Hec05}, (4) \citealt{Mag98},, (5) \citealt{Gre03b}, (6) \citealt{Pet04}, 
	  (7) \citealt{Lod03}, (8) \citealt{Den10}, (9) \citealt{Den09}, (10) \citealt{Ben06}, (11) \citealt{Gu06}, (12) \citealt{Bia07}, (13) \citealt{Mat99}, 
	  (14) \citealt{Bia05a}, (15) \citealt{Gre08}, (16) \citealt{Gil10}, (17) \citealt{Cre12},
	  (18) \citealt{Shu07}, (19) \citealt{Ris99}, (20) \citealt{Mat04a}, (21) \citealt{Bia05b}, (22) \citealt{Mar09}, (23) \citealt{Mat04b}, 
	  (24) \citealt{Ris02}, (25) \citealt{Gua04}, (26) This work, (27) \citealt{Deo09}, (28) \citealt{Ves06}, (29) \citealt{Gru04}, 
	  (30) \citealt{Mul06}, (31) \citealt{Mul09}, (32)\citealt{Hic09}}
\label{observables}
\end{deluxetable}
\end{landscape}

\begin{deluxetable}{lrrl}
%\tabletypesize{\scriptsize}
\tablecolumns{2}
\tablecaption{Correlations between modeled/observed AGN properties}
\tablewidth{0pt}
\tablehead{\colhead{Correlation}  & \colhead{$r$} & \colhead{$N$} & \colhead{$P_{c}(r,N)$} }
\startdata
$i$ vs $N_H$                 & .86 & 15 & 0.000039\\
$z_{max}$ vs $r_{t}$         & .82 & 16 & 0.00010 \\
$i$ vs IR Color\tablenotemark{1}              &-.74 & 15 & 0.0016  \\
$i$ vs H$\beta$ FWHM         & .94 &  5 & 0.017   \\
$r_{t}$ vs L$_{Bol}$         & .56 & 16 & 0.024   \\
$v_{max}$ vs L$_{Bol}$       & .52 & 16 & 0.039   \\
$z_{max}$ vs L$_{Bol}$       & .43 & 16 & 0.096   \\
$\beta$ angle vs $N_H$       &-.33 & 16 & 0.21    \\
$v_{max}$ vs $r_{t}$         & .32 & 16 & 0.23    \\
$v_{max}$ vs $z_{max}$       & .31 & 16 & 0.24    \\
$v_{max}$ vs $M_{H_2}$       & .53 &  4 & 0.47    \\
$\theta_{max}$ vs L$_{Bol}$  & .17 & 16 & 0.53    \\
$v_{max}$ vs $\theta_{max}$  &-.13 & 16 & 0.63    \\
$\theta_{max}$ vs $M_{H_2}$  & .25 &  4 & 0.75    \\

\enddata
\tablenotetext{1}{$F_{5.5{\mu}m}/F_{30{\mu}m}$}
\label{coefficients}
\end{deluxetable}

\clearpage

\bibliographystyle{apj}             % Please learn to use the
                                     %  formatting of Latex's Bibtex. It
                                     %  will make your life easier.
% apj.bst should be in this directory as well as apj-jour.bib and reference paper.bib
\bibliography{apj-jour,paper2}       % "paper.bib" contains all my
                                     %  references. "apj-jour.bib"
                                     %  contains abbreviations of
                                     %  journals.
% to get references to work in paper:
% compile paper       --> latex mrk78.tex
% compile bibtex      --> bibtex mrk78
% compile paper again --> latex mrk78.tex

\clearpage

%FIGURES%%%%%%%%%%%%%%%%%%%%%%%%%%%%%%%%%%%%%%%%%%%%

% Gauss fits to spectra

%\appendices
%\include{appendix-slitpos}
%\include{appendix-specimg}
%\include{appendix-kin}
%\endappendices

\end{document}